\documentclass{article}

 \usepackage[final]{neurips_2025}

\usepackage[utf8]{inputenc} % allow utf-8 input
\usepackage[T1]{fontenc}    % use 8-bit T1 fonts
\usepackage{hyperref}       % hyperlinks
\usepackage{url}            % simple URL typesetting
\usepackage{booktabs}       % professional-quality tables
\usepackage{amsfonts}       % blackboard math symbols
\usepackage{nicefrac}       % compact symbols for 1/2, etc.
\usepackage{microtype}      % microtypography
\usepackage{xcolor}  % colors
\usepackage[table]{xcolor}
\usepackage{amsmath} 

\setcitestyle{square,sort,comma,numbers}
\usepackage{graphicx}
\usepackage{caption}
\usepackage{subcaption}

\title{Natural gradient descent for improving variational inference based classification of radio galaxies}

\author{%
  Devina Mohan \\
  Department of Physics \& Astronomy\\
  University of Manchester, UK\\
  \texttt{devina.mohan@postgrad.manchester.ac.uk} \\
  \And
  Anna Scaife\thanks{The Alan Turing Institute, 96 Euston Rd, London, UK \texttt{a.scaife@turing.ac.uk}} \\
  Department of Physics \& Astronomy\\
  University of Manchester, UK\\
  \texttt{anna.scaife@manchester.ac.uk} \\
}

\begin{document}

\maketitle
\begin{abstract}
Bayesian neural networks (BNNs) are most commonly optimised with first-order optimisers such as stochastic gradient descent. 
However, when optimising for parameters of probabilistic models, incorporating second order information during optimisation can lead to a more direct path in the distribution space and faster convergence.
In this work we examine whether using natural gradient descent can improve the performance of variational inference based classification of radio galaxies. We use the Improved Variational Online Newton (iVON) algorithm and compare its performance against a recent benchmark for BNNs for radio galaxy classification. We find that iVON results in better uncertainty calibration out of all the methods previously considered while providing similar predictive performance to the best performing inference methods such as Hamiltonian Monte Carlo and Bayes by Backprop based variational inference. Models trained with iVON can distinguish far out-of-distribution optical galaxy data, but they cannot reliably detect radio galaxy images from a telescope with different resolution and sensitivity. We find that the cold posterior effect persists in the models trained with iVON. 
Our results suggest that the choice of the optimiser can lead to qualitatively different solutions and future work using probabilistic neural network models should carefully consider the inductive biases being encoded through the optimisation process, in addition to the data, architecture and inference method.

\end{abstract}

% Please read the instructions below carefully and follow them faithfully. \textbf{Important:} This year the checklist will be submitted separately from the main paper in OpenReview, please review it well ahead of the submission deadline: \url{https://neurips.cc/public/guides/PaperChecklist}.

\section{Introduction}

Upcoming radio astronomy surveys are expected to produce exabyte scale data which  requires statistically robust machine learning models to extract scientific information \cite{an2019science}. Bayesian neural networks (BNNs) provide a principled way to model uncertainty in deep learning models. A recent benchmark of Bayesian deep learning for radio galaxy classification suggests that Hamiltonian Monte Carlo \citep[HMC; ][]{neal2011mcmc, cobb2021scaling} performs well in terms of predictive accuracy, calibration of uncertainty and detecting distribution shift
\citep{pmlr-v244-mohan24a}. However, given the computational cost of HMC, the authors suggest that improving variational inference \citep[VI; ][]{kingma2015variational, practicalvi} is a promising direction for future radio surveys. The authors use the Bayes by Backprop \citep[BBB-VI; ][]{blundell} algorithm to implement variational inference which performs reasonably well compared to all the other approximations considered but faces some challenges such as sensitivity to initialisation, slow convergence and the cold posterior effect (according to which the posterior must be down-weighted to get good predictive performance). 

% why do we expect second order optimisation to be better for variational inference than SGD?
In variational inference,  the negative Evidence Lower Bound (ELBO) is most commonly optimised using Stochastic Gradient Descent (SGD) based updates of the variational parameters.
% which uses local first-order information about the loss function. 
Gradient descent algorithms assume that the parameters can be treated as Cartesian coordinates in a Euclidean space, which is a reasonable assumption for standard point-wise neural networks. % and makes updates based on Euclidean geometry of the weight space. 
However, since VI is optimising a family of distributions over the neural network parameters, the parameter space forms a statistical manifold where every point on the manifold corresponds to a probability distribution over some domain $\mathcal{X}$. 
%The field of information geometry studies statistical manifolds \cite{amari1998natural}. 
Distances on such manifolds are measured by defining a metric that induces an inner product on the tangent space at each point of the manifold. The space of all possible distributions of neural network parameters is a Riemannian manifold, with the Fisher information matrix (FIM) as its metric \citep{amari2019fisher}. 
% It can also be shown that there exist two Riemannian manifolds that each capture the functional change and the reparameterisation properties of the parameters \cite{roy2025reparameterization}. 
Therefore to account for the geometry of the variational parameters, it is possible to precondition the gradient with the inverse of the FIM. This technique is known as natural gradient descent (NGD).
% , makes the optimisation well conditioned.
% need to link second order optimisation with NGD.
NGD can also be viewed through the lens of second order optimisation if the FIM is used to approximate the Hessian. 
% Thus accounting for the geometry of the variational parameters through second order information about the loss function might provide a better conditioned objective for optimising the probability distributions being inferred by VI. 

In this work we examine whether using NGD
can improve variational inference based classification of radio galaxies. Although the FIM is intractable to calculate for large deep learning models, early work shows how adding adaptive weight noise to natural gradient for point estimation can be used to approximate natural gradient variational inference updates \cite[Noisy Natural Gradient;][]{zhang2018noisy}. %explain limitations and how is it different than ivon
Recent advances in the field of deep learning optimisation have led to the development of scalable variational learning algorithms like Improved Variational Online Newton based on natural gradient descent \citep[iVON;][]{ivon, blr_emti, pmlr-v119-lin20d}. We use the iVON optimiser and find that the posterior predictive distributions from iVON provide better calibrated uncertainties than all the approximate Bayesian inference methods considered in the previous benchmark %in terms of the mean UCE but the still within 1sigma
, while providing similar predictive performance to the best performing models. However, this comes at the cost of reduced ability to detect distribution shift from near out-of-distribution (OoD) radio data, while preserving the ability to detect far-OoD data like images of optical galaxies.

\section{Natural Gradient Descent for Variational Inference with iVON}

Instead of optimising for point-wise model parameters, $\boldsymbol{\theta}$, with maximum likelihood or maximum a posteriori estimation (in case regularisation is used), a Bayesian neural network places a prior over the model parameters, $p(\boldsymbol{\theta})$, and infers a posterior probability distribution $ p(\boldsymbol{\theta}|\mathcal{D}) $ over the model parameters given data, $\mathcal{D}$,
% = {(x_i, y_i)}_{i=1}^n$
using Bayes rule. Since this posterior is intractable to compute for neural networks, various approximations are used \citep{blundell, gal2015bayesian, cobb2021scaling, daxberger2021laplace}. VI is one such approximation in which the posterior $p(\boldsymbol{\theta}|\mathcal{D})$ is approximated with a parametric distribution, $q(\boldsymbol{\theta})$, from a family of distributions $\mathcal{Q}$
% also known as the variational distribution 
\citep{vireview}. To find the optimal parameters of this variational distribution, the negative ELBO cost function is optimised:
% with gradient descent:  
%
\begin{equation}
    \mathcal{L}(q) \equiv  \min_{q(\boldsymbol{\theta}) \in \mathcal{Q}} \,  - \mathbb{E}_{q(\boldsymbol{\theta})}[ \log p(\mathcal{D}|\boldsymbol{\theta})] + \,\mathrm{KL}[q(\boldsymbol{\theta}) || p(\boldsymbol{\theta})], 
    \label{eq:elbo_t}
\end{equation}
where $p(\mathcal{D}|\boldsymbol{\theta})$ is the likelihood. 
The improved Variational Online Newton (iVON) algorithm builds on the Bayesian Learning Rule \cite[BLR;][]{blr_emti} for minimal exponential families. Many learning algorithms can be derived as a special case of the Bayesian Learning Rule (BLR) which optimises the following variational objective to approximate the posterior:
% using a parametric distribution $q(\theta)$. 
%
\begin{equation}
    \mathcal{L}(q) \equiv  
    % \min_{q(\theta) \in \mathcal{Q}} \,  \mathbb{E}_{q(\theta)}[l(\theta)] + \mathrm{KL}[q(\theta) || p(\theta)] \equiv 
    \min_{q(\boldsymbol{\theta}) \in \mathcal{Q}} \, \mathbb{E}_{q(\boldsymbol{\theta})}[l(\boldsymbol{\theta})] - \mathcal{H}(q). 
    \label{eq:blr}
\end{equation}
Here Equation \ref{eq:elbo_t} is re-written in terms of a generic loss function defined as $l(\boldsymbol{\theta}) = -\log p(\mathcal{D}|\boldsymbol{\theta}) \, p({\boldsymbol{\theta}})$ and the 
% which is usually the negative log likelihood, $-p(D|\theta)$, and the KL divergence term is expressed in terms of
entropy, $\mathcal{H}(q)$. 
% The temperature term is subsumed in a later step of the optimisation algorithm, see Equation \ref{eq:ivon_update}. 
If the variational distribution is chosen to be a minimal exponential family with natural parameters $\boldsymbol{\lambda}$ and expectation parameters $\boldsymbol{\mu}$, the above objective can be minimised by exploiting the information geometry of the posterior using natural gradients to update the natural parameters at current optimisation step, $t$, as follows \cite{khan2025information}:
% to find the best candidate using the following update rule:
\begin{equation}
    \boldsymbol{\lambda}_{t+1} \leftarrow \boldsymbol{\lambda}_t - \alpha \, \mathbf{F}^{-1} \nabla_{\boldsymbol{\lambda}} \mathcal{L}(\boldsymbol{\lambda}) , 
    \label{eq:natural_grad}
\end{equation}
where $\alpha$ is the learning rate
% , $\hat{g} = $ is the natural gradient 
and $\mathbf{F}$ is the FIM of the natural parameters. 
% $\mathbf{F}:= - \mathbb{E}_q[\nabla^2_{\boldsymbol{\lambda}} \log q(\boldsymbol{\theta}|\boldsymbol{\lambda})]$. 
The bijection between the natural and expectation parameter space allows one to write the natural gradient with respect to the natural parameters as a gradient with respect to the expectation parameter and the following update is derived \cite{blr_emti}:
% Here, the variational objective is written in terms of the natural parameters. 
\begin{equation}
    \boldsymbol{\lambda}_{t+1} \leftarrow \boldsymbol{\lambda}_t - \alpha \nabla_{\boldsymbol{\mu}} \{\mathbb{E}_{q_{{\boldsymbol{\lambda}}_t}} [l(\boldsymbol{\theta})] - \mathcal{H}(q) \}.
    \label{eq: blr_update}
\end{equation}
%
% where $\boldsymbol{\lambda}$ is the natural parameter, and the gradient is taken with respect to the expectation parameter $\boldsymbol{\mu}$. 
% For minimal exponential families, the gradient with respect to the expectation parameter can be written as a natural gradient with respect to the natural parameter, $\boldsymbol{\lambda}$. The gradient of the entropy wrt to the natural parameter is just the natural parameter: $\nabla_m \mathcal{H}(q) = \boldsymbol{\lambda}$
% Here the natural parameters is updated by calculating gradients with respect to the expectation parameters. 
In the following equations, we will denote $\mathbb{E}_{q_{{\boldsymbol{\lambda}}_t}}$ as $\mathbb{E}_{q_{_t}}$ for clarity.
By making different choices of the variational distribution $q(\boldsymbol{\theta})$ and different approximations to the natural gradient, one can derive gradient descent, Newton's method, Stochastic VI and many other learning algorithms \citep{blr_emti}.

% iVON makes a Gaussian approximation over network weights. 
% minimal exponential family
When $q(\boldsymbol{\theta})$ is chosen to be a multivariate Gaussian, $q(\boldsymbol{\theta}) = \mathcal{N}(\boldsymbol{\theta}|\mathbf{m}, \mathbf{S}^{-1})$,
% with mean $\mathbf{m}$ and precision $\mathbf{S^{-1}}$, 
% \begin{equation}
% \end{equation}
BLR resembles Newton's update: $
    \boldsymbol{\theta} \leftarrow \boldsymbol{\theta}  - \mathbf{H}_{\boldsymbol{\theta}}[\nabla_{\boldsymbol{\theta}} l(\boldsymbol{\theta})]$, 
where $\mathbf{H} = \nabla^2_{\boldsymbol{\theta}} l(\boldsymbol{\theta})$ is the hessian of the loss.
For a multivariate Gaussian, the natural parameters in terms of the mean $\mathbf{m}$ and precision $\mathbf{S}$ are $\boldsymbol{\lambda} := \{\mathbf{S}\mathbf{m}, -\mathbf{S}/2\}$ and the expectation parameters are $\boldsymbol{\mu}:= \{ \mathbb{E}_q(\boldsymbol{\theta}), \mathbb{E}_q (\boldsymbol{\theta} {\boldsymbol{\theta}}^T)\}$. Plugging the values of the first and second natural and expectation parameters in Equation \ref{eq: blr_update}
% and applying the Bonnet's and Price's theorem
, the following update rules can be derived for the mean and the precision in terms of the gradient and Hessian of the loss with samples from $q(\boldsymbol{\theta})$ \cite{blr_emti}: %mean and correlation 
% \begin{equation}
%     \boldsymbol{\theta} \leftarrow \boldsymbol{\theta}  - \mathbf{H}_\boldsymbol{\theta}[\nabla_\boldsymbol{\theta} l(\boldsymbol{\theta})]
% \end{equation}
%
% \begin{equation}
%     \boldsymbol{\lambda} \leftarrow \boldsymbol{\lambda} - \rho \nabla_\boldsymbol{\mu} \mathbb{E}_q[l(\boldsymbol{\theta}) - \mathcal{H}(q)]
%     \label{}
% \end{equation}
%mean update
\begin{equation}
    \mathbf{m}_{t+1} \leftarrow \mathbf{m}_t - \alpha_t \mathbf{S}^{-1}_{t+1} \mathbb{E}_{q_t} [\nabla_{\boldsymbol{\theta}} l(\boldsymbol{\theta})] 
    \label{eq: mean_update}
\end{equation}
%precision update
\begin{equation}
    \mathbf{S}_{t+1} \leftarrow (1-\alpha_t) \mathbf{S}_t + \alpha_t\mathbb{E}_{q_t}[\nabla^2_{\boldsymbol{\theta}} l(\boldsymbol{\theta})]
    \label{eq:precision_update}
\end{equation}
% BLR performs natural gradient variational inference, but under a constrained parameterisation $\boldsymbol{\lambda}$.

The precision update involves the hessian which is intractable to compute. 
% An earlier version of the algorithm used a Gauss-Newton estimate to approximate the Hessian which requires computing per sample squared gradients. 
% To reduce the computational cost, the Hessian estimate is calculated using the reparameterisation trick %The precision is also used to scale the mean update
% iVON adapts the optmisation step size for every parameter based on the local curvature of the loss function. 
The iVON algorithm makes further simplifications to scale the learning algorithm to large models and datasets. 
It makes a diagonal approximation to the hessian and uses the reparameterisation trick to calculate second order information using the mini-batch gradients 
%
% \begin{equation}
$    \mathbf{\hat{g}} = \hat{{\nabla}} l (\boldsymbol{\theta}),$ $\;  \boldsymbol{\theta} \sim \mathcal{N}(\boldsymbol{\theta} | \mathbf{m}, \textrm{diag} (\boldsymbol{\sigma})):
$
% \end{equation}
% explain a little about how the reparameterisation trick is used to approximat the FIM/Hessian. 
%
\begin{equation}
    \mathbf{\hat{h}} = \mathbf{\hat{g}} \: (\boldsymbol{\theta} - \mathbf{m})/ \boldsymbol{\sigma}^2.
    \label{eq:hess_ivon}
\end{equation}
Thus iVON approximates second order curvature information by measuring how first order information (gradient) is affected by random perturbations of the parameters. 
An exponentially moving average of the gradient and hessian estimates is used to update the mean and standard deviation values. 
Assuming a multivariate Gaussian variational distribution 
% $\mathcal{N}(\mathbf{\theta}|\boldsymbol{\mu}, \mathbf{S})$
implies that the precision matrix $\mathbf{S}$ must be positive-definite \cite{pmlr-v119-lin20d}. To handle the positive-definite constraint, an additional geometric correction term derived from approximate Riemannian gradient descent is added to the Hessian approximation, $\mathbf{h}$, to ensure its positivity:
\begin{equation}
    \mathbf{h} \leftarrow (1-\rho) \mathbf{h}+ \rho \mathbf{\hat{h}} + \frac{1}{2} \rho^2 (\mathbf{h} - \mathbf{\hat{h}})^2 / (\mathbf{h} + s_0/\mathrm{ess}),
\end{equation}
where $s_0$ is the prior precision assuming $p(\boldsymbol{\theta}) = \mathcal{N}(\boldsymbol{\theta}|0, \mathbf{I}/s_0)$, $\mathrm{ess}$ is the effective sample size (ESS), and $\rho > 0$ is a constant. In practice, the weight decay regulariser, $\delta$, is used instead of prior precision. Setting the ESS to the size of the training set, $\mathrm{ess} = N$,
% _{\mathrm{train}}$
is equivalent to setting the temperature of the posterior to 1. Setting $\mathrm{ess} > N$ gives cold posteriors which may be required in practice to get good predictive performance \cite{wenzel2020good, mohan2022quantifying, pmlr-v244-mohan24a}.
% o control the temperature of the posterior, which is known to be important for achieving good performance [16, 17, 4], an effective sample size (ESS) parameter that directly scales the learned precision h. 
% Setting ess = N corresponds to a standard variational posterior (T=1), while setting ess > N produces colder, more confident posteriors.
% = s_0/\textrm{ess}.
The geometric correction term 
% arises from performing the optimization on the Riemannian manifold of probability distributions. The standard update for the variance can lead to invalid (non-positive) values, equivalent to stepping 'off the manifold'. This correction term 
% adjusts the update to follow the curvature of the space, 
mathematically guarantees that the estimated precision remains positive and the variational distribution remains valid throughout training.
The mean and standard deviation of the model parameters are then calculated as follows:
\begin{equation}
    \mathbf{m} \leftarrow \mathbf{m} - \alpha \, \frac{ (\mathbf{\hat{g}} + \delta \mathbf{m})}{(\mathbf{h} + \delta)} ; \; \, \,\,\,\,\,\,\,\,\, 
    \boldsymbol{\sigma} \leftarrow \frac{1}{\sqrt{\mathrm{ess} (\mathbf{h} + \delta)}}.
    \label{eq:ivon_update}
\end{equation}
%
% where $\mathrm{ess}$ is the effective sample size (ESS) and $\delta$ is the weight decay, which is used instead of specifying a prior.
For the mean estimation, the numerator $(\mathbf{\hat{g}} + \delta \mathbf{m})$ gives the direction of update and the denominator $(\mathbf{h} + \delta)$ gives an adaptive estimate of the curvature for each parameter. The authors provide a PyTorch implementation of iVON which facilitates its use similar to standard first-order optimisation algorithms like Adam. 
The updates in Equation \ref{eq:ivon_update} can be contrasted with the BBB-VI updates which calculate separate gradients of the variational objective with respect to the variational parameters in the Euclidean space given learning rate $\alpha$: $ \mathbf{m} \leftarrow \mathbf{m} - \alpha \nabla_{\mathbf{m}}\mathcal{L}(\boldsymbol{\theta})$ and $\boldsymbol{\sigma} \leftarrow \boldsymbol{\sigma} - \alpha\nabla_{\boldsymbol{\sigma}}\mathcal{L}(\boldsymbol{\theta})$.
% \textbf{Linearised Laplace Approximation}
% % % How is the Hessian being approximated in case of LLA?

% \citep{roy2025reparameterization}

% Explain reparameterisation invariance.
% \citep{roy2025reparameterization} recently showed that 

\section{Experimental Setup}

\textbf{Data and architecture}
We use the MiraBest Confident dataset \citep{porter2023mirabest} which consists of Fanaroff-Riley type I (FR1) and type II (FRII) radio galaxies to train a LeNet like architecture as specified in the previous benchmark \cite{pmlr-v244-mohan24a}.
Additionally, we use the MIGHTEE \cite{heywood2022mightee} and GalaxyMNIST\footnote{\url{https://github.com/mwalmsley/galaxy_mnist}} datasets to evaluate the ability of the method to detect distribution shift. Details are recalled in the appendix for completeness.

\textbf{iVON experiments}
We use the iVON optimiser \citep{ivon} to train the model for classification. 
% We find that the initial learning rate needs to be set much higher than typical supervised learning problems.
We set the initial learning rate to $0.2$ and use a cosine annealing based learning rate scheduler to anneal the learning rate to 0 after a warm up period of 5 epochs. The optimisation process is sensitive to the initialisation of the hessian which is specified through the $h_0$ hyperparameter. We test different values of $h_0 = \{0.01, 0.1, 0.5, 1, 5\}$ and find the best results are achieved with $h_0 = 0.5$.
The effective sample size (ESS) hyperparameter, $\mathrm{ess}$, is used to set the temperature of the posterior.  We experiment with different values of ESS, $\textrm{ess} = \{0.1N, N, 10N, 100N \}$ and find that using a cold posterior with $\textrm{ess} = 10N$ gives the best performance in terms of predictive performance and calibration of uncertainty. 
The weight decay is set to $\delta = 10^{-4}$, which is the same value used in the SGD based BBB-VI implementation of \cite{pmlr-v244-mohan24a}. To train the model we use one Monte Carlo sample as recommended in the original implementation of iVON, which is also consistent with the BBB implementation of VI used in the previous benchmark. Results are reported for models trained for 1000 epochs using 10 different random seeds and random shuffling of data between training and validation samples. We use an early stopping criterion to select the model with the lowest validation loss during training. The epoch at which early stopping comes into effect is different for each of the 10 runs, ranging from epoch 189 to epoch 972. Training a single model takes $25$ mins on an NVIDIA A100 GPU.
% \textcolor{red}{The authors write that the model was trained for 1000 epochs but use early stopping. How many epochs were required until early stopping kicked in?}
% with different ESS values -- for example ESS = len data = temp 1 - converges much more slowly and performs worse
Code for this work is available at \url{ https://github.com/devinamhn/RadioGalaxies-BNNs}

In addition to experiments with iVON, we reproduce BBB-VI and HMC results from the previous benchmark for Bayesian deep learning for radio galaxy classification from \cite{pmlr-v244-mohan24a} for comparison.

\section{Results}

We construct the posterior predictive distribution with $200$ Monte Carlo samples and calculate the predictive accuracy, Uncertainty Calibration Error (UCE) and energy scores for detecting distribution shift. UCE is a measure of the \textit{inherent} calibration of the Bayesian posterior predictive distribution, which we calculate using using the 64\% credible intervals of the distributions. 
% We note here that BNNs do not have frequentist coverage guarantees.
Evaluation metrics are recalled in Appendix \ref{apd:metrics} for completeness. 

Experimental results for iVON are reported in Table \ref{tab:eval} for two different ESS values, along with results reproduced from the previous benchmark for BBB-VI and HMC from \cite{pmlr-v244-mohan24a}. 
% Works with much higher initial learning rate $0.2$
We find that iVON with $\mathrm{ess} = 10N$ has similar predictive performance to BBB-VI and better UCE than BBB-VI, but it converges much faster (25mins vs 40 mins). Setting  $\mathrm{ess} = 100N$ achieves similar predictive performance and uncertainty calibration to BBB-VI. 

% explain UCE metric here 'inherent calibration of the posterior posterior predictive distribution'
% \textcolor{red}{evaluation metric on uncertainty can be mixing frequentist notion and Bayesian notion --- a  perfectly calculated posterior might still not have nominal frequentist coverage e.g., having a very informative but not quite correct prior. It is okay to evaluate overall frequentist properties like the authors do, but it is good to be clear what was been tested.} 
 
\begin{figure}
     \centering
      \begin{subfigure}[b]{0.49\textwidth}
         \centering
         \includegraphics[width=\linewidth]
         {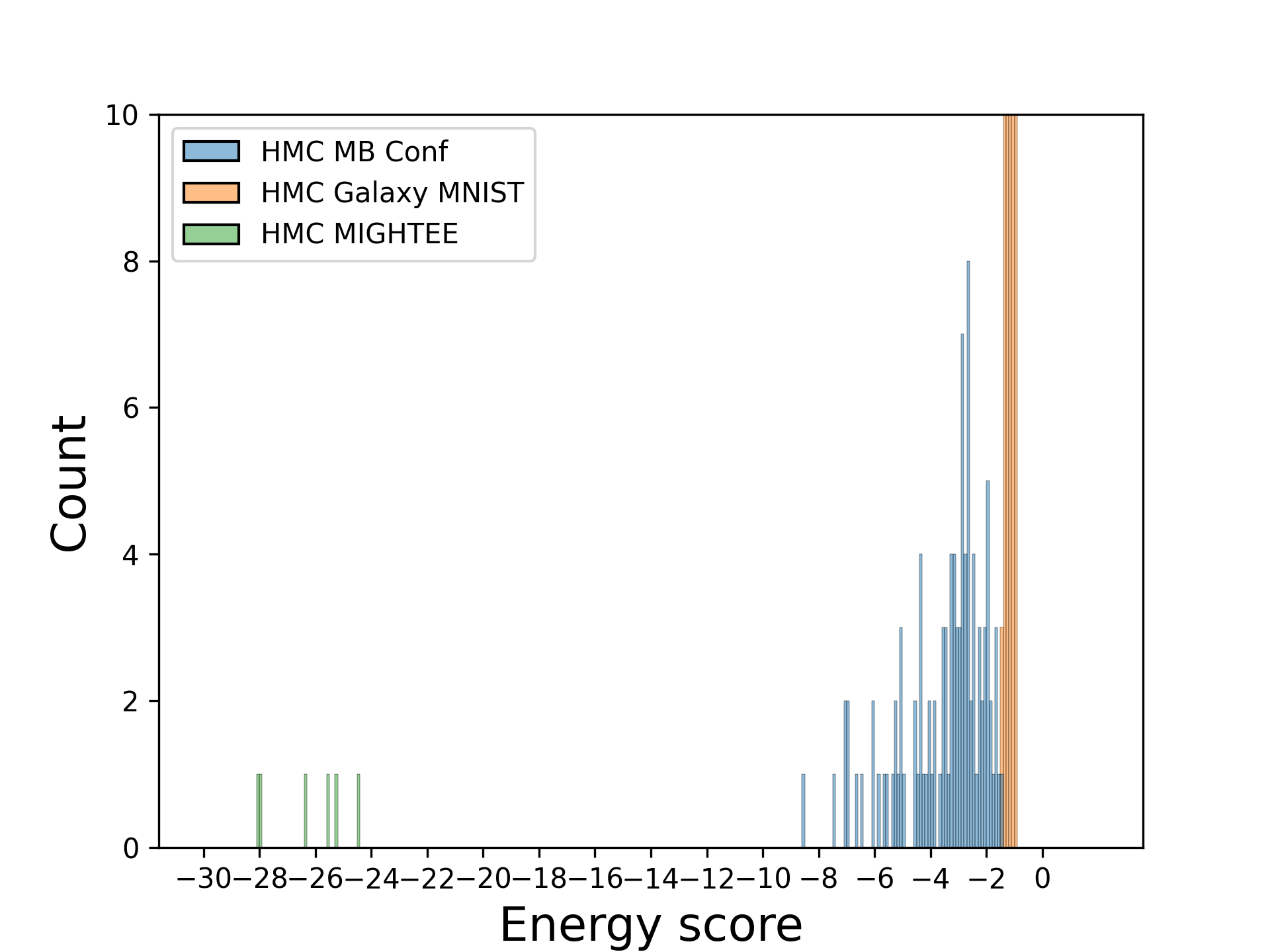}
         \caption{}
         \label{fig:hmc_energy}
     \end{subfigure}
     \hfill
     \begin{subfigure}[b] {0.49\textwidth}
         \centering
         \includegraphics[width=\linewidth]
         {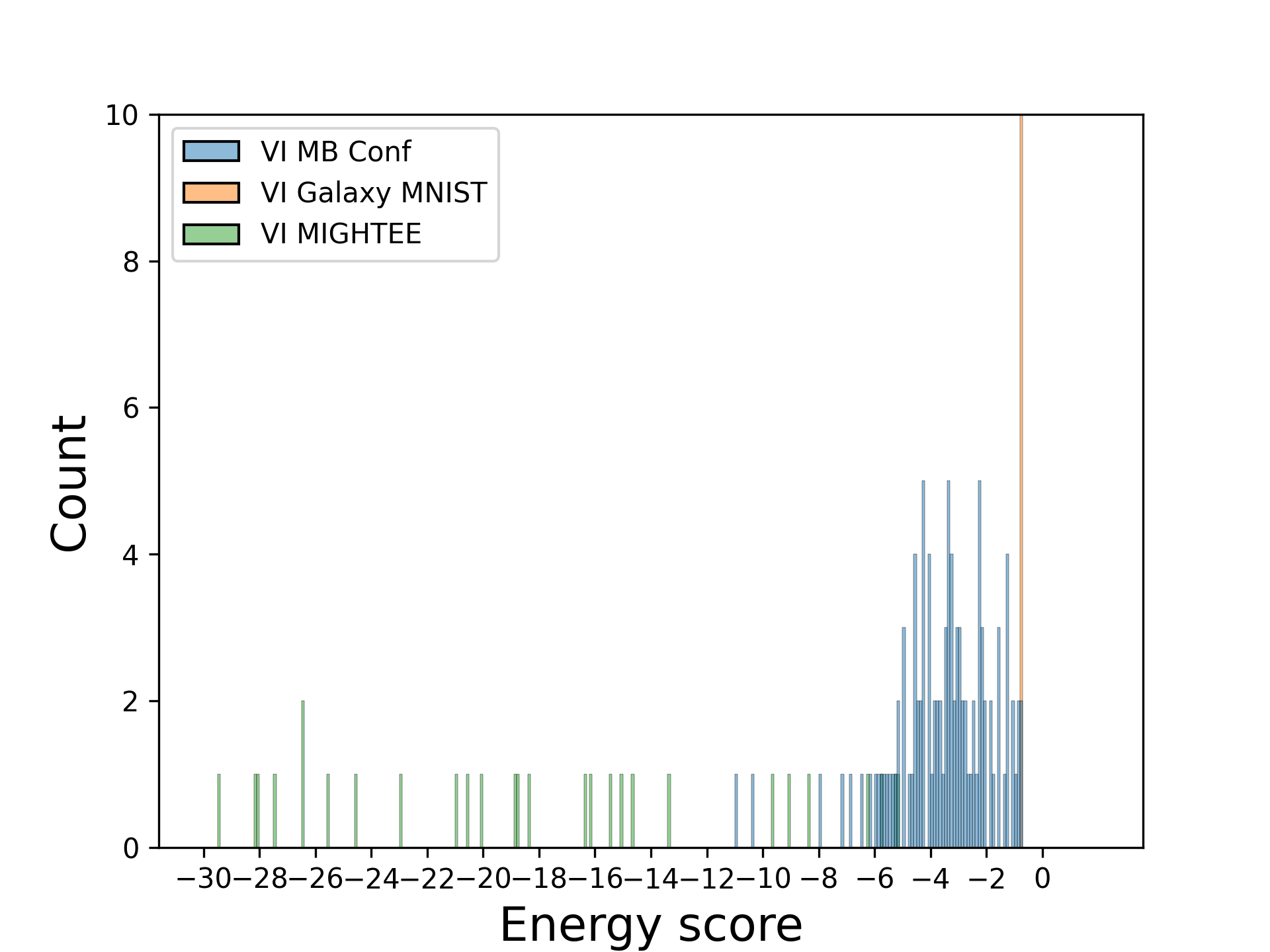}
         \caption{}
         \label{fig:bbb_energy}
     \end{subfigure}
     \hfill
     \begin{subfigure}[b]{0.49\textwidth}
         \centering
         \includegraphics[width=\linewidth]
         {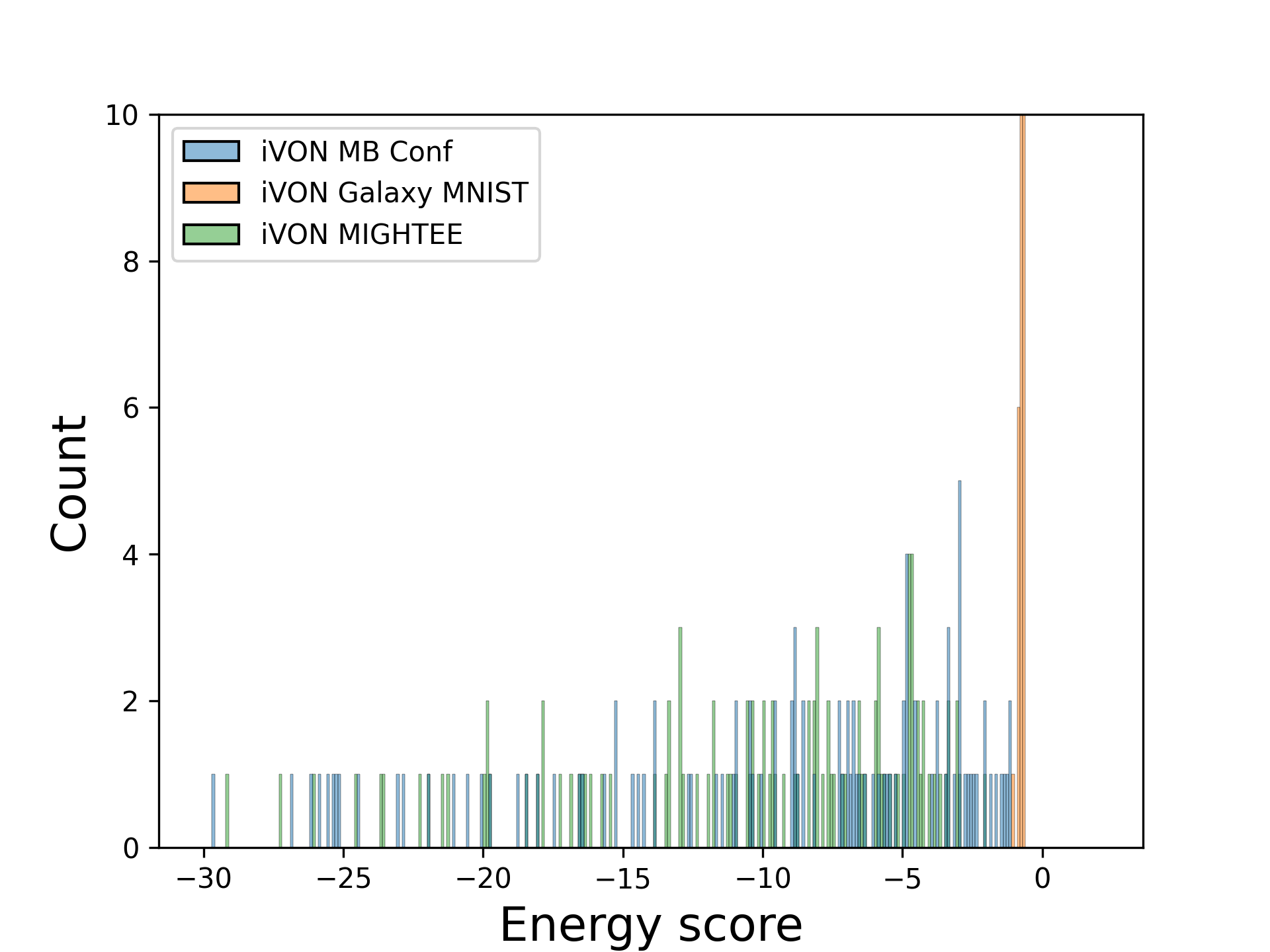}
         \caption{}
         \label{fig:ivon_energy}
     \end{subfigure}
        \caption{ Histograms of energy scores calculated for the MiraBest (MBConf; blue), GalaxyMNIST (orange) and MIGHTEE (green) test datasets for (a) HMC (b) BBB-VI and (c) iVON. The HMC and BBB-VI plots are reproduced from \cite{pmlr-v244-mohan24a}.}
        \label{fig:energy}
\end{figure}

We plot the histograms of energy scores for the different test sets in Figure\ref{fig:energy} to examine the ability of iVON to detect distribution shift. Out-of-distribution (OoD) data is expected to have higher energy scores. We use the model with the lowest validation error from the experiments to calculate the energy scores. Energy plots for BBB-VI and HMC are reproduced from the previous benchmark \cite{pmlr-v244-mohan24a} for comparison. We find that while iVON is able to clearly distinguish optical galaxy images, it cannot reliably detect distribution shifted radio galaxies from the MIGHTEE dataset compared to BBB-VI, see Figures \ref{fig:bbb_energy} and \ref{fig:ivon_energy}. HMC inference is still the most reliable in detecting distribution shift from different radio galaxy datasets, see Figure \ref{fig:hmc_energy}. One might expect the Mirabest test set to have the lowest energy score since the model was trained on radio galaxies from the same dataset, however, MIGHTEE has even lower energy scores. This is because MIGHTEE contains images of radio galaxies from a telescope with different resolution and sensitivity but the types of objects in the dataset are similar to the training data (i.e, FRI and FRII galaxies). So these objects are \textit{near-OoD}, rather than \textit{far-OoD} like the optical galaxies from GalaxyMNIST which have completely different morphological features and have a more significant distribution shift. Near-OoD detection is a more challenging problem than detecting far OoD data.
% Discuss \cite{li2025position}.
% "Covariate shifts are label preserving, and can include common noise corruptions and transformations, or going from hand-written to type-written characters, or from natural "im%https://arxiv.org/pdf/2007.08176

% \textcolor{red}
% {Regarding the energy scores, I expected that the training data set has the lowest energy score since it should be iD (training data). However, MIGHTEE has the lowest energy score. Why is this the case?}

% Still requires tempering through the 'effective sample size (ess)' term.
%
\begin{table*}[!ht]
    \centering
    \caption{Test error, uncertainty calibration error (UCE) of the predictive entropy and time to train for HMC, BBB-VI and iVON based Bayesian neural networks. The HMC and BBB-VI results are reproduced from \cite{pmlr-v244-mohan24a}. For iVON, results are shown for two different values of the effective sample size (ess) as a multiple of the length of training data, $N$.
    % We also provide a baseline MAP error percentage. 
    % Inference methods with a ($^*$) indicate that no data augmentation was used during inference for those experiments. - removed data augmentation results, iVON does not work without data augmentation at all 
    % See Sections \ref{sec:predictive performance} and \ref{sec:Uncertainty Calibration}.
    } 
    \label{tab:eval}
    \begin{tabular}{rlll}
        \toprule % from booktabs package
        \bfseries Inference  & \bfseries Error (\%) $\downarrow$  & \bfseries UCE $\downarrow$ & \bfseries Time \\
        \midrule % from booktabs package
        HMC & $4.16 \pm 0.45 $   & $14.76 \pm 0.95$ & 7 days \\
        % HMC$^{*}$ & $6.24 \pm 0.45$ & $12.65 \pm 0.01$  \\
        BBB-VI & $3.94 \pm 0.01 $  & $12.77 \pm 6.11$ & 40 mins\\
        iVON $\textrm{ess}=10N$ & $3.07 \pm 1.47 $   & $8.37 \pm 4.12 $ & 25 mins\\ %exp6
        iVON $\textrm{ess}=100N$& $3.36 \pm 1.23 $   & $12.19 \pm 6.57 $ & 25 mins \\ %exp3
        % VI-BBB$^{*}$  & $3.84 \pm 0.01 $  & $12.32 \pm 6.36$\\
        % VI (linear conditioner) &  & &\\
        % LLA (diag) & $6.86 \pm 0.17$ & & 22.39\\
        % LLA (Laplace library) & $8.85 \pm 2.09$  & $23.84 \pm 3.54$\\
        % \rowcolor{yellow} Linearized Laplace (diag GGN from posteriors) & $ 8.26\pm 1.83 $  & $ 23.87\pm 1.35$\\
        % \rowcolor{yellow} Linearized Laplace (EF) & $ \pm $  & $ \pm $\\
        % Dropout &  $7.88 \pm  2.81$ &  $25.75 \pm 4.44$ \\     
        % Ensembles & $7.69 \pm 0.27 $  & 24.41 \\
        % MAP & $5.76$  & \\
        % \bottomrule 
    \end{tabular}
\end{table*}

\newpage

\section{Conclusions and Discussion}

% discuss reparameterisation invariance in natural gradient descent, geometry etc 
We find that accounting for the geometry of the variational parameters through second order information provides a better conditioned objective for optimising the probability distributions being inferred by VI. 
% discuss optimiser inductive bias
For overparameterised models, different optimisers can converge to different optimal solutions while minimising the same objective function. 
% Implicit regularisation occurs when the dynamics of optimisation produce regularisation. 
Our results suggest that optimising the variational objective with iVON, which is a natural gradient descent based optimiser, provides better uncertainty calibration than SGD optimisation, and improved convergence speed while providing the same predictive performance. However, the ability to distinguish between different radio galaxy datasets is reduced compared to BBB-VI. 
This suggests that even if an optimiser converges to a reasonable solution faster, as is the case when we use NGD instead of SGD to optimise VI, it might not find a solution that will be good for a diverse range of objectives. 
% while HMC is able to cleanly distinguish between different datasets. 
Different optimisers induce different inductive biases \cite{pascanu2025optimizers}. They constrain the space of solutions that is reachable by any optimiser within the space of all solutions defined by the data, architecture, initialisation, which can lead to different types of representations (distributed and redundant vs localised and compressed, for example). Future work could look at formalising these trade-offs to induce the most useful inductive biases for a particular application domain.

% However, VI will always find only a single mode. 
% HMC is able to cleanly distinguish between different datasets while local approximations like VI will always find only a single mode.

% Recent work has shown that there exist two Riemannian manifolds that each capture the functional change and the reparameterisation properties of the parameters of Bayesian neural networks \cite{roy2025reparameterization}.  

% gradient descent trains faster in directions of large feature variance (top principle components) and then overfits to the noise unless regularisation is deployed. This encodes an inductive bias which can hurt the generalisation performance when correalations between different dimensions might contain useful information.
% does robustness to ood require longer training times so that the low curvature directions can also be trained sufficiently? Maybe add more regularisation to ivon and train it for longer? 

% iVON posterior predictive is better calibrated than HMC

% Do {\bf not} include this section in the anonymized submission, only in the final paper. You can use the \texttt{ack} environment provided in the style file to autmoatically hide this section in the anonymized submission.

\begin{ack}
AMS gratefully acknowledges support from an Alan Turing Institute AI Fellowship EP/V030302/1.
\end{ack}

% \section*{References}
% \clearpage
\bibliography{references}

\clearpage

%%%%%%%%%%%%%%%%%%%%%%%%%%%%%%%%%%%%%%%%%%%%%%%%%%%%%%%%%%%%

\appendix

\section{Data}
Radio galaxies are characterised by large scale jets and lobes which can extend up to mega-parsec distances from the central black hole and are observed in the radio spectrum. The original binary classification scheme proposed to classify such extended radio sources was based on the ratio of the extent of the highest surface brightness regions to the total extent of the galaxy \citep{fr1974}. Fanaroff Riley Type I (FRI) galaxies are edge-darkened whereas Fanaroff Riley Type II (FRII) galaxies are edge-brightened. Over the years, several other morphologies such as bent-tail \citep{rudnick1976, odea1985owen}, hybrid \citep{gopalkrishna2000}, and double-double \citep{schoenmakers2000} sources have also been observed and there is still a continuing debate about the exact interplay between extrinsic effects, such as the interaction between the jet and the environment, and intrinsic effects, such as differences in central engines and accretion modes, that give rise to the different morphologies. 

We train our BNNs on the MiraBest Confident dataset [Section \ref{sec:mirabest}] and use the MIGHTEE [Section \ref{sec:mightee}] and GalaxyMNIST [Section \ref{sec:galaxy_mnist}] datasets to test the ability of our BNNs to detect different types of distribution shifts.

\subsection{MiraBest Dataset}
\label{sec:mirabest}
The MiraBest dataset used in this work consists of 1256 images of radio galaxies of $150\times 150$ pixels pre-processed to be used specifically for deep learning tasks \citep{porter2023mirabest}. The galaxies are labelled using the FRI and FRII morphological types based on the definition of \citep{fanaroff1974morphology} and further divided into their subtypes. In addition to labelling the sources as FRI, FRII and their subtypes, each source is also flagged as `Confident' or `Uncertain' to indicate the human classifiers' confidence while labelling the dataset. In this work we use the MiraBest Confident subset and consider only the binary FRI/FRII classification. The training and validation sets are created by splitting the predefined training data into a ratio of 80:20. The final split consists of 584 training samples, 145 validation samples, and 104 withheld test samples. Data augmentation in the form of random rotations is used.

\subsection{MIGHTEE}
\label{sec:mightee}
%domain adaptation
The MIGHTEE dataset is constructed using the Early Science data products from the MeerKAT International GHz Tiered Extragalactic Exploration survey \citep[MIGHTEE;][]{heywood2022mightee}. MIGHTEE is an ongoing radio continuum survey being conducted using the MeerKAT telescope, which is one of the precursors to the Square Kilometer Array (SKA). The survey provides radio continuum, spectral line and polarisation data, of which we use the radio continuum data and extract images for the COSMOS and XMMLSS fields. While there are thousands of objects in these fields, expert labels are only available for $117$ objects. We use the data pre-processing and expert labels made available by \cite{slijepcevic2024radio}. The dataset contains classifications based on the consensus of five expert radio astronomers. The final sample contains $45$ FRI and $72$ FRII galaxies
% , see Figure \ref{fig:data} (second row).
We note that the MIGHTEE dataset contains significant observational differences from the MiraBest dataset.

\subsection{Galaxy MNIST}
\label{sec:galaxy_mnist}
% Optical images - completely out-of-distribution. 

In addition to considering different datasets of radio galaxies which have been curated using using data from radio telescopes, we also evaluate our models on data collected from optical telescopes. 
%diference in radio vs optical images
Optical images of galaxies contain different features and in a sense represent completely out-of-distribution galaxies which well-calibrated models should classify with a very high degree of uncertainty so that they can be flagged for inspection by an expert.
We use the GalaxyMNIST\footnote{\url{https://github.com/mwalmsley/galaxy_mnist}} dataset which contain images of $10,000$ optical galaxies classified into four morphological types using labels collected by the Galaxy Zoo citizen science project
% , see Figure \ref{fig:data} (last row) for examples. 
The galaxies are drawn from the Galaxy Zoo Decals catalogue \citep{walmsley2022galaxy}. We resize the high resolution images from 224x224 to 150x150 to match the input dimensions of our model. We construct a small test set of $104$ galaxies from the dataset to evaluate the out-of-distribution detection ability of our BNNs.

\section{Evaluation metrics}
\label{apd:metrics}

We use the expected value of the posterior predictive distribution to obtain the classification of each galaxy in the MiraBest test set and calculate the test error for a single experimental run by taking an average of the classification error over the entire test set. We report the mean and standard deviation of the test error for $10$ experimental runs. 
% , see Table \ref{tab:eval}.

\subsection{Predictive entropy}

Using Monte Carlo (MC) samples obtained from the posterior predictive distributions of different BNNs, one can obtain $N$ Softmax probabilities for each class, $c$, in the dataset. %In practice we make $N$ forward passes through the network to get the samples. - in implementation
We can recover $N$ class-wise Softmax probabilities as follows:
\begin{equation}
    P(y|x, D) = \frac{1}{N}\sum_{i=1}^N P(y=c|x, w^{(i)}) ,
    \label{eq:softmax_probs}
\end{equation}
where $w^{(i)}$ is the $i^{\textrm{th}}$ weight sample from the posterior distribution conditioned on the training data, $D$, and $(x, y)$ are data from the test set. %We denote this quantity as $S_c$ to simplify the notation in the following sections.
Using these samples, one can quantify the uncertainty in the predictions using different metrics. In this work we look at the predictive entropy of the softmax distribution which measures the average amount of information inherent in the distribution and is defined as:
\begin{equation}
     %\mathbb{H}(y|x, D) = - \sum_c q(y=c|x, \theta^{(i)}) \log q(y=c|x, \theta^{(i)}) ,
     \mathbb{H}(y|x, D) = - \sum_c P(y=c|x, w) \log P(y=c|x, w) ,
\end{equation}
which can be approximated using MC samples as follows \citep{gal2016uncertainty}:
\begin{equation}
    \mathbb{H}(y|x, D) = - \sum_c \left (\frac{1}{N} \sum_{i=1}^N P(y=c|x, w^{(i)}) \right) \log \left (\frac{1}{N} \sum_{i=1}^N P(y=c|x, w^{(i)}) \right) .
    \label{eq:pred_entropy}
\end{equation}
%

%We use the natural logarithm for all the equations described in this section and the values are reported in nats, which is the natural unit of information. The entropy thus attains a maximum value of $\sim 0.693$ nats, when the predictive entropy is maximum and a minimum value close to zero.

\subsection{Uncertainty Calibration Error (UCE)}

We report the expected uncertainty calibration error \cite[UCE;][]{gal2015bayesian, laves2019well, mohan2022quantifying} of the predictive entropy of our posterior predictive distributions in Table \ref{tab:eval}. We use the $64\%$ credible intervals of the posterior predictive distributions to calculate UCE. To examine how well calibrated the predictive entropy is, we calculate the Uncertainty Calibration Error (UCE), which is a more general form of the Expected Calibration Error (ECE). UCE is a weighted average of the difference between fractional error and uncertainty calculated for the output of the model when binned into $M$ bins of equal width for a particular uncertainty metric:
\begin{equation}
    \textrm{UCE} = \sum_{m=1}^M \frac{|B(m)|}{n}| \textrm{err}(B_m) - \textrm{uncert}(B_m)|.
\end{equation}
Here $B_m$ is the set of data in a particular bin, $n$ is the total number of data points and $\textrm{ uncert}(B_m)$ is the average value of a given uncertainty metric for those data points:
%
%Uncertainty metrics are normalised between 0 and 1 and partitioned into $M$ bins. The normalised uncertainty values are then used to calculate the average uncertainty in bin $B_m$ as follows:
%
\begin{equation}
    \textrm{uncert}(B_m) = \frac{1}{|B_m|} \sum_{i \in B_m} \textrm{uncert}_i ~ ,
\end{equation}
where $\textrm{uncert}_i$ can be calculated using Equation~\ref{eq:pred_entropy} followed by minmax-normalisation to bring values into the range 0 to 1. 

Equation 16 of \cite{laves2019well} defines the average fractional error in bin $B_m$ to be:
\begin{equation}
 \textrm{err}(B_m) = \frac{1}{|B_m|} \sum_{i \in B_m} \textrm{err}_i~,
\end{equation}
where $\textrm{err}_i$ is the contribution to this error from an individual data point, defined as:
\begin{equation}
 \textrm{err}_i = \mathbf{1} (\hat{y}_i \neq y)~~~~~\forall~~i \in B_m~.
\end{equation}
Here we redefine $\textrm{err}_i$ to be the average error obtained for an individual data sample, such that
\begin{equation}
 \textrm{err}_i = \frac{1}{N} \sum_{j=1}^N \mathbf{1} (\hat{y}_{ij} \neq y)~~~~~\forall~~i \in B_m~,
\end{equation}
where $N$ is the number of samples drawn from the posterior predictive distribution.

\subsection{Energy score}
\cite{liu2020energy} propose a post-hoc scoring function for discriminative classification models which can be used to distinguish between in-distribution (iD) and distribution-shifted data examples. We calculate average scalar energy scores for different test samples, $x$, for all the datasets considered in this work using the logit values, $f_i(x)$, for each class, $i$, using  $N$ posterior samples:
\begin{equation}
    \tilde{\mathrm{E}}(x; f) = \frac{1}{N}  \sum_j^{N} \, -T . \,  \textrm{log} \sum_i^K e^{f_{i} (x) / T } ,
\end{equation}
where the temperature term, $T$, is set to $1$. Out-of-distribution (OoD) samples are expected to have higher energy in this framework.

\end{document}